\documentstyle[prd,aps,epsfig]{revtex}
\begin{document}
\title{Non-factorizable effects in the $B-\bar B$ mixing} 
\author{
Dmitri Melikhov$^{a,b}$ and Nikolai Nikitin$^{b}$}
\address{
$^a$ ITP, Universit\"at Heidelberg, Philosophenweg 16, D-69120, Heidelberg, 
Germany\\
$^b$ Nuclear Physics Institute, Moscow State University, 
119899, Moscow, Russia}
\maketitle
\begin{abstract} 
We analyse the corrections to factorization for the $B-\bar B$ mixing amplitude 
assuming that non-factorizable soft gluon exchanges can be described by the 
local gluon condensate. 
Within this approximation the 
$\langle\alpha_s GG\rangle$-correction to the mixing amplitude can be expressed 
through the $B$-meson transition form factors at zero momentum transfer. 
The correction is {\it negative} independent of the particular
values of the form factors. 

As a next step, we study these form factors making use of  
the relativistic dispersion approach based on the constituent quark picture. 
We obtain spectral representations for the form factors in terms 
of the $B$-meson soft wave function and the propagator of the color-octet 
$b\bar q$ system. The $B$-meson  wave function has been determined previously 
from weak exclusive $B$ decays, so the main uncertainty in our estimates comes 
from the unknown details of the propagator in the nonperturbative region. 
We obtain $\Delta B_{B_d}(m_b)=-0.06\pm 0.035 $ and 
$\Delta B_{B_s}(m_b)=-0.05\pm 0.03 $. 
\end{abstract}
\section{Introduction}
The study of the oscillations in the system of neutral $B$ mesons provides important 
information on the pattern of the CP violation in the Standard model and its extentions 
(a detailed discussion can be found in \cite{buras}). 
The main uncertainty in the theoretical description of the process arises from  
the nonperturbative long-distance contributions to the mixing amplitude.  
In the $B-\bar B$ mixing there are two kinds of such contributions:  
first, effects related to the presence of the $B$-mesons in the initial and final states, 
and, second, corrections to the weak 4-quark amplitude due to the soft gluon exchanges 
between quarks of the initial $B$ and the final $\bar B$ meson. 
By neglecting the second contribution the amplitude factorizes.   
In this case all nonperturbative effects connected with the meson formation are reduced 
to only one quantity - the leptonic decay constant $f_B$. 

Soft-gluon corrections to the weak 4-quark amplitude 
give rise to non-factorizable contributions, such that the influence of the 
$B$-meson structure is no longer described by the decay constant only. 
The understanding of the actual size of the non-factorizable effects and their 
adequate description in the heavy quark systems is an important and challenging task. 

The theoretical analysis of $B-\bar B$ mixing was first performed within 
the QCD sum rules \cite{sr1,sr2,sr3} and later using lattice QCD, see \cite{lat1,lat2} and 
refs therein. The results from the sum rules are in all cases well compatible with 
factorization (vacuum saturation), within rather large errors. 
Recent lattice analyses definitely reported negative non-factorizable effects, 
around 5-10\% for the central values at the scale $\mu\simeq m_b$.  
Still, the errors of the calculations have a similar size. 
Results from these approaches are shown in Table \ref{table:results}. 
\begin{table}[htb]
\caption{\label{table:results}
Comparison of results from various approaches}
\centering
\begin{tabular}{|l|l|l|l|l|}
 Ref.     &$B_{B_d}(m_b)$&$B_{B_s}(m_b)$&$\hat B_{B_d}$&$\hat B_{B_s}/\hat B_{B_d}$\\
\hline
SR\cite{sr1,sr2}&  $1.0\pm 0.15$	&		&	       &	      \\
SR  \cite{sr3}  &  $0.95\pm 0.1$	&		&	       &	      \\
\hline
Lat \cite{lat1} &0.92(4)$^{+3}_{-0}$&0.91(2)$^{+3}_{-0}$&  1.41(6)$^{+5}_{-0}$&0.98(3)\\
Lat \cite{lat2} &   0.93(8)    &   0.92(6)    &   1.38(11)   &  0.98(5)     \\
\hline
This work  &   $0.94\pm 0.035$  & $0.95\pm 0.03$ & $1.4\pm 0.05$ & $1.01\pm 0.01$    
\end{tabular}
\end{table}

In this letter we analyse corrections to factorization in the $B-\bar B$ mixing 
due to soft gluon exchanges in the leading $\alpha_s$-order, assuming   
that the main effect of such exchanges can be described by the local 
gluon condensate \cite{svz}. 
We show that within this approximation the parameter 
$\Delta B_B$ which describes correction to factorization, is determined by the meson 
transition form factors at zero momentum transfer. It is important that 
$\Delta B_B$ turns out to be {\it negative}, independently of the particular values 
of these form factors. 
 
As a next step we calculate the relevant form factors. 
To this end we make use of the relativistic dispersion approach based on the 
constituent quark picture \cite{m}. 
We obtain spectral representations for the form factors in terms of the $B$-meson 
wave function and the propagator of the color-octet $b\bar q$-system. 

\newpage  
Parameters of the model, such as the quark masses and the wave functions, are known 
from the analysis of the weak $B$-meson form factors \cite{mb,ms} within the same 
dispersion approach. For numerical estimates we also need the propagator of the 
color-octet $q\bar b$-system in the nonperturbative region, so we discuss the possible 
form of this propagator. The lack of reliable information about its details in the 
nonperturbative region leads to the main uncertainties in our numerical estimates. 

The paper is organised as follows: 
The next Section presents basic formulas for the $B-\bar B$ mixing. In
Section III we calculate the non-factorizable contribution to the mixing amplitude 
induced by the local gluon condensate and show that the correction is negative. 
In Section IV we calculate the $B$-meson transition form 
factors within the dispersion approach. Section V gives numerical estimates. 
\section{Effective Hamiltonian and the structure of the amplitude}
\begin{figure}[htb]
\begin{center}
\mbox{\epsfig{file=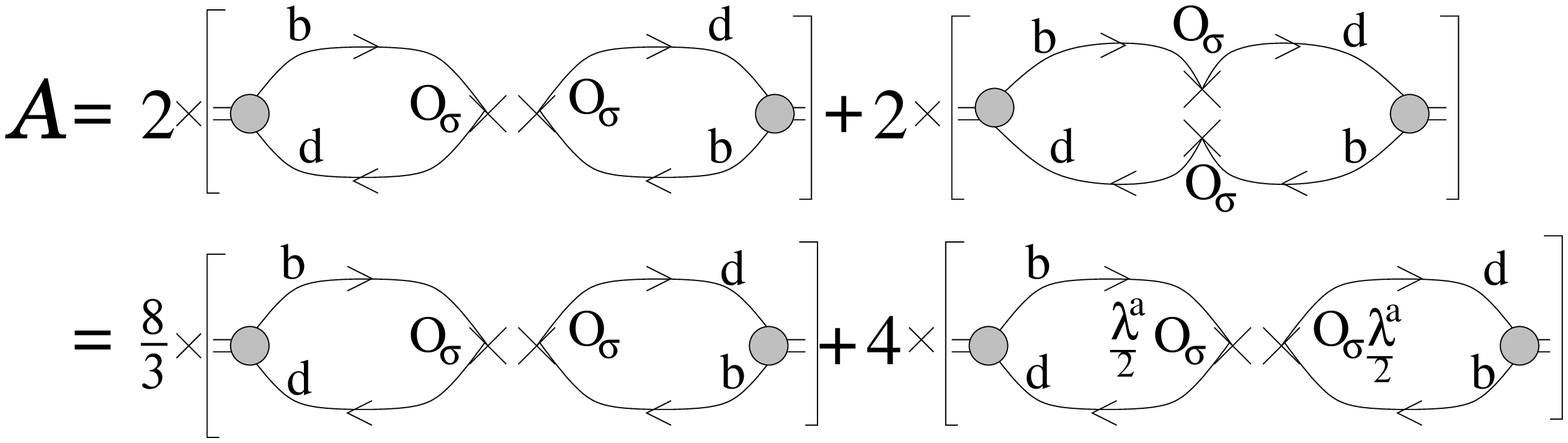,height=3cm}}
\caption{\label{fig:fig1}
Different representations for the $B^0-\bar B^0$ mixing amplitude. 
The enclosed quark lines show the order of contraction of the spinorial indices. 
The circles stand for the initial (final) $B$ ($\bar B$) meson. The second line 
is obtained from the first line by performing the combined color and spinorial 
Fierz rearrangements.}
\end{center}
\end{figure}

The effective Hamiltonian which summarizes the perturbative QCD corrections for the 
$B^0-\bar B^0$ mixing has the following form \cite{heff}
\begin{equation}
\label{heff1}
H^{\Delta B=2}_{eff}=\frac{G^2_F\, M^2_W}{\sqrt{2}}\,
\left (V^*_{tb}V_{td}\right )^2\,C(\mu)\,
\left (\bar b O_{\sigma} d\right )\left (\bar b O_{\sigma} d\right ) + h.c.
\end{equation}
where $G_F$ is the Fermi constant, $O_{\sigma}=\gamma_{\sigma}(1-\gamma_5)$,
$\gamma^5=i\gamma^0\gamma^1\gamma^2\gamma^3$. 
The parameter $\mu$ stands for the renormalization scale which separates the 
hard region 
taken into account perturbatively and the soft region. 
The Wilson coefficient $C(\mu)$ includes perturbative corrections above the scale $\mu$.
The explicit expression for $C(\mu)$ can be 
taken from \cite{heff}. The soft contributions are contained in the 
$B$-meson matrix elements of the operators in the effective Hamiltonian 
\begin{eqnarray}
\label{me1}
A=\left<\bar B^0 \left |\left (\bar b O_{\sigma} d \right )
\left (\bar b O_{\sigma} d \right )\right | B^0\right>.  
\end{eqnarray}
The diagrammatic representation for the amplitude $A$ of Eq. (\ref{me1}) is shown in 
Fig \ref{fig:fig1}. 
Using the language of the hadronic intermediate states one finds that the 
contribution of the hadronic vacuum to this amplitude gives 
\begin{equation}
\label{vac}
\left < \bar B^0 | H^{\Delta B=2}_{eff} | B^0 \right > =
\frac{8}{3}\,\frac{G^2_F\, M^2_W\, M^2_B}{\sqrt{2}}\,
(V^*_{tb}V_{td})^2\, C(\mu)\, f^2_B, 
\end{equation}
with $f_B$ the leptonic decay constant of the $B$ meson defined according to the relation  
$\left < 0 |\bar b\gamma_{\mu}\gamma_5 d | B(p)\right > = i\,f_B\,p_{\mu}$.
It is convenient to parametrize the full amplitude as follows 
\begin{equation}
\label{hadr}
\left < \bar B^0 | H^{\Delta B=2}_{eff} | B^0 \right > =
\frac{8}{3}\,\frac{G^2_F\, M^2_W\, M^2_B}{\sqrt{2}}\,
(V^*_{tb}V_{td})^2\, C(\mu)\, f^2_B\, B_B,
\end{equation}
such that the quantity $B_B-1\equiv \Delta B_B$  
measures contributions of the non-vacuum intermediate hadronic states. 

Using the language of quarks and gluons, the corrections to the amplitude of Fig. 1 due to 
soft gluon exchanges 
are obtained by inserting the (soft) gluons between the quark lines 
in Fig 1. 
Soft gluon exchanges between the quarks of the same loop 
lead to the $\alpha_s$-corrections either to the meson vertices or to the quark propagators. 
They only contribute to the leptonic decay constant $f_B$ but do not lead to 
non-factorizable effects. 

\begin{figure}[tb]
\begin{center}
\begin{tabular}{c}
\mbox{\epsfig{file=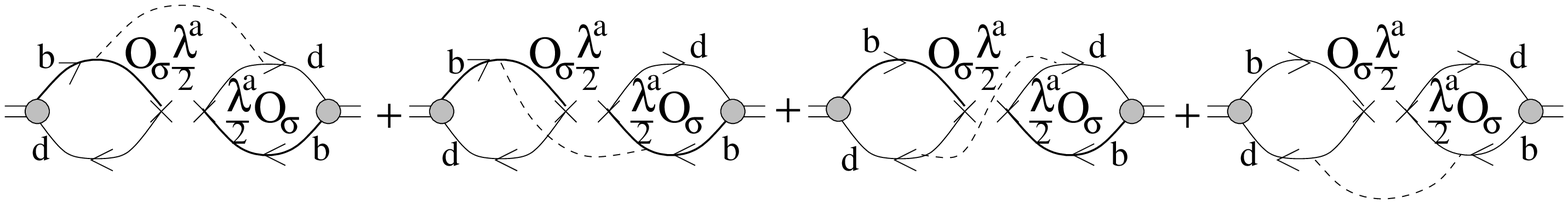,width=16cm}}
\end{tabular}
\caption{\label{fig:fig2} Soft-gluon exchanges leading to the non-factorizable effects.}
\end{center}
\end{figure}

The non-factorizable effects originate from the soft gluon exchanges between 
quarks of different loops. To lowest $\alpha_s$-order these effects are described by the 
4 diagrams shown in Fig. \ref{fig:fig2}. 
We assume that the main effect of the soft gluon exchange can be described by
the local gluon condensate. In the next section we demonstrate that 
in this approximation the $\left<\alpha_sGG\right>$-correction to the factorization 
is {\it negative}. 
\section{$\Delta B$ in terms of the local gluon condensate} 
\begin{figure}[tb]
\begin{center}
\begin{tabular}{cc}
\mbox{\epsfig{file=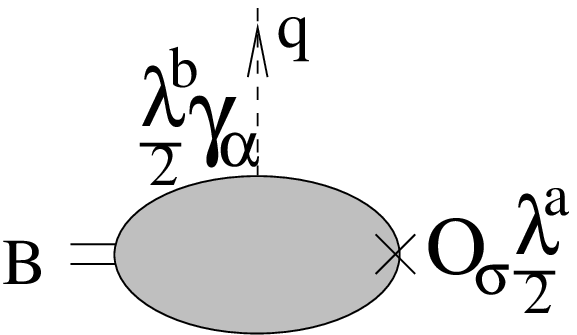,width=3cm}}&\qquad\mbox{\epsfig{file=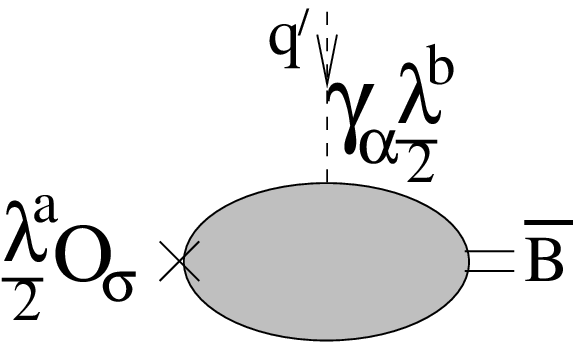,width=3cm}}
\end{tabular}
\caption{\label{fig:3} Diagrams for 
$T^{ab(l)}_{\sigma\alpha}(p,q)$ (a) and 
$T^{ab(r)}_{\sigma\alpha}(p,q)$ (b).}
\end{center}
\end{figure}

Let us consider the exchange of a soft gluon between different quark loops 
assuming the dominance of the local gluon condensate \cite{svz}. 
In this case a typical graph of Fig 2 describing the soft-gluon contribution is reduced 
to the product of the three-point diagrams as shown in Fig \ref{fig:3}. 
The corresponding amplitude has the form
\begin{eqnarray}
\label{aa1}
A^{(1)}=g^2
\int dq dq' T^{ab(l)}_{\sigma\alpha}(p,q)
            T^{ab'(r)}_{\sigma\alpha'}(p,q)
\frac{dx}{(2\pi)^4} \frac{dx'}{(2\pi)^4} {\rm e}^{-iqx+iq'x'}\langle A^b_\alpha(x)A^{b'}_{\alpha'}(x')\rangle,   
\end{eqnarray}
where $T^{ab(l)}_{\sigma\alpha}(p,q)$ with the superscript $l$ stands for the amplitude  
of Fig \ref{fig:3}a with the $B$ in the initial state,  
and $T^{ab'(r)}_{\sigma\alpha'}(p,q')$ with the superscript $r$ for the amplitude of Fig 
\ref{fig:3}b with the $\bar B$ in the final state.  

It is convenient to use the fixed-point gauge $x_\mu A^b_\mu(x)=0$ \cite{srff}. In this case 
the gluon potential reads 
\begin{eqnarray}
\label{gauge}
A^b_\alpha(x)=\frac{1}{2}x_\beta G^b_{\alpha\beta}+...,
\end{eqnarray}
where dots stand for terms with higher order derivatives.  
The average over the hadronic vacuum is performed according to the relation 
\begin{eqnarray}
\langle G^b_{\alpha\beta}G^{b'}_{\alpha'\beta'}\rangle=\frac{\delta^{bb'}}{96}
(g_{\alpha\alpha'}g_{\beta\beta'}-g_{\alpha\beta'}g_{\alpha'\beta})\langle GG \rangle,  
\end{eqnarray}
with $\delta^{aa}=N_c^2-1$. Here $\langle GG \rangle$ is the positive-valued gluon condensate 
$\langle\frac{\alpha_s}{\pi} GG \rangle=0.012\;{\rm GeV}^4$ \cite{svz}. 

The amplitude of Fig 3 then takes the form 
\begin{eqnarray}
\label{aa2}
A^{(1)}&=&\frac{g^2}{4}
\int dq dq'  
T^{ab(l)}_{\sigma\alpha}(p,q)   \frac{\partial}{\partial q_\beta}\delta(q)
T^{ab'(r)}_{\sigma\alpha'}(p,q')\frac{\partial}{\partial q'_{\beta'}}\delta(q') 
\langle G^b_{\alpha\beta} G^{b'}_{\alpha'\beta'}\rangle
\nonumber\\
&=&
\frac{g^2}{4}\langle G^b_{\alpha\beta} G^{b'}_{\alpha'\beta'}\rangle
\frac{\delta^{ab }}{2\sqrt{N_c}}T^{(l)}_{\sigma\alpha, \beta }(p)
\frac{\delta^{ab'}}{2\sqrt{N_c}}T^{(r)}_{\sigma\alpha',\beta'}(p), 
\end{eqnarray}
where we have denoted 
\begin{eqnarray}
\label{relation}
\frac{\partial}{\partial q_\beta}T^{ab(l,r)}_{\sigma\alpha}(p,q)|_{q=0}\equiv
\frac{\delta^{ab}}{2\sqrt{N_c}}T^{(l,r)}_{\sigma\alpha,\beta}(p). 
\end{eqnarray}
Notice that only the parts of the amplitudes 
$T^{(l,r)}$ antisymmetric in indices $\alpha$ and $\beta$ give a nonvanishing contribution. 
These antisymmetric parts have the following general Lorentz structure: 
\begin{eqnarray}
\label{aaa} 
T^{A(l,q)}_{\sigma\alpha,\beta}&=&
4p^\nu\left[f_1^q \epsilon_{\alpha\beta\sigma\nu}
+i f_2^q (g_{\alpha\nu}g_{\beta\sigma}-g_{\beta\nu}g_{\alpha\sigma})\right], 
\nonumber\\
T^{A(l,\bar q)}_{\sigma\alpha,\beta}&=&
4p^\nu\left[f_1^{\bar q} \epsilon_{\alpha\beta\sigma\nu}
+i f_2^{\bar q} (g_{\alpha\nu}g_{\beta\sigma}-g_{\beta\nu}g_{\alpha\sigma})\right],
\nonumber\\ 
T^{A(r,q)}_{\sigma\alpha,\beta}&=&T^{A(l,\bar q)}_{\sigma\alpha,\beta}, \nonumber\\
T^{A(r,\bar q)}_{\sigma\alpha,\beta}&=&T^{A(l,q)}_{\sigma\alpha,\beta}, 
\nonumber\\ 
&&f_1^q=f_1^{\bar q}, \qquad f_2^q=-f_2^{\bar q}.  
\end{eqnarray}
The superscript $q(\bar q)$ in these expressions denotes the flavour of the 
quark (antiquark) to which the soft external gluon is attached in diagrams of Fig 3.  
The quantities $f_{1,2}^{q,\bar q}$ are real constants. The last three lines 
in Eq. (\ref{aaa}) are the consequence of the $C$-invariance of 
the strong interaction $S$-matrix. As we discuss in the next section, 
the constants $f_{1,2}^{q,\bar q}$ can be represented as specific $B$-meson 
transition form factors at zero momentum transfer.   

We have to collect now contributions of the four diagrams in Fig 2. 
For instance, the contribution to the amplitude of the subprocess in which the soft gluons 
are attached to the $b$ quark in the left loop and the $\bar b$ quark in the right loop 
reads 
\begin{eqnarray}
\label{correction}
T^{A(l,b)}_{\sigma\alpha;\beta}(p)T^{A(r,\bar b)}_{\sigma\alpha';\beta'}(p)
(g_{\alpha\alpha'}g_{\beta\beta'}-g_{\alpha\beta'}g_{\alpha'\beta})
=-192 M_B^2 \left((f^b_1)^2+(f^b_2)^2\right). 
\end{eqnarray}
Similar expressions are easily obtained for other diagrams using the relations (\ref{aaa}). 
Taking into account the relevant symmetry factors as shown in Fig 1,  
we finally arrive at the following expressions (we also list the factorizable amplitude 
in the same normalization)  
\begin{eqnarray}
\label{a1}
A^{(0)}&\simeq&\frac{8}{3}M_B^2f_B^2\nonumber\\
A^{(1)}&\simeq& -8C_F M_B^2 \left< \frac{\alpha_s}{\pi}GG\right> \pi^2
\left[(f_1^{b}+f_1^{d})^2+(f_2^{b}-f_2^{d})^2
\right], 
\end{eqnarray}
with the color factor $C_F=\frac{N_c^2-1}{4N_c}$. 
Finally, for $\Delta B_B$ we find the expression 
\begin{eqnarray}
\label{deltab}
\Delta B_B(\mu)=-\frac{\left< \frac{\alpha_s}{\pi}GG\right>}{f_B^2}2\pi^2
\left[
\left(f_1^{b}+f_1^{d}\right)^2+\left(f_2^{b}-f_2^{d}\right)^2
\right]. 
\end{eqnarray}
Obviously, the correction to the factorizable amplitude due to the 
local gluon condensate is {\it negative}. This agrees with all numerical results for 
$\Delta B$. Notice that the constants $f_1$ and $f_2$ depend on the 
renormalization scale $\mu$. 
\section{Form factors in the dispersion approach} 
We now proceed to the calcuation of the form  factors $f_1$ and $f_2$.  
To this end we make use of the relativistic dispersion approach based on the 
constituent quark picture. Within this approach all observables are given by the spectral 
representations in terms of the soft wave function of the participating mesons. 

\begin{figure}[htb]
\begin{center}
\begin{tabular}{lr}
\mbox{\epsfig{file=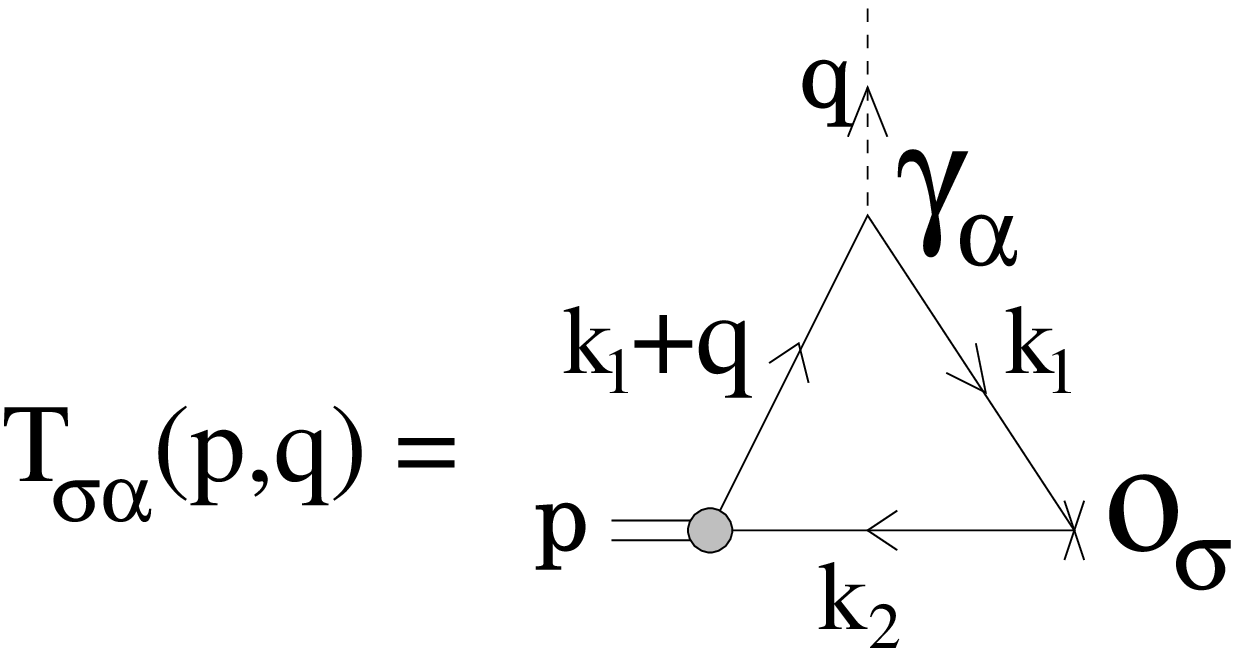,width=7.cm}} & $\qquad$
\mbox{\epsfig{file=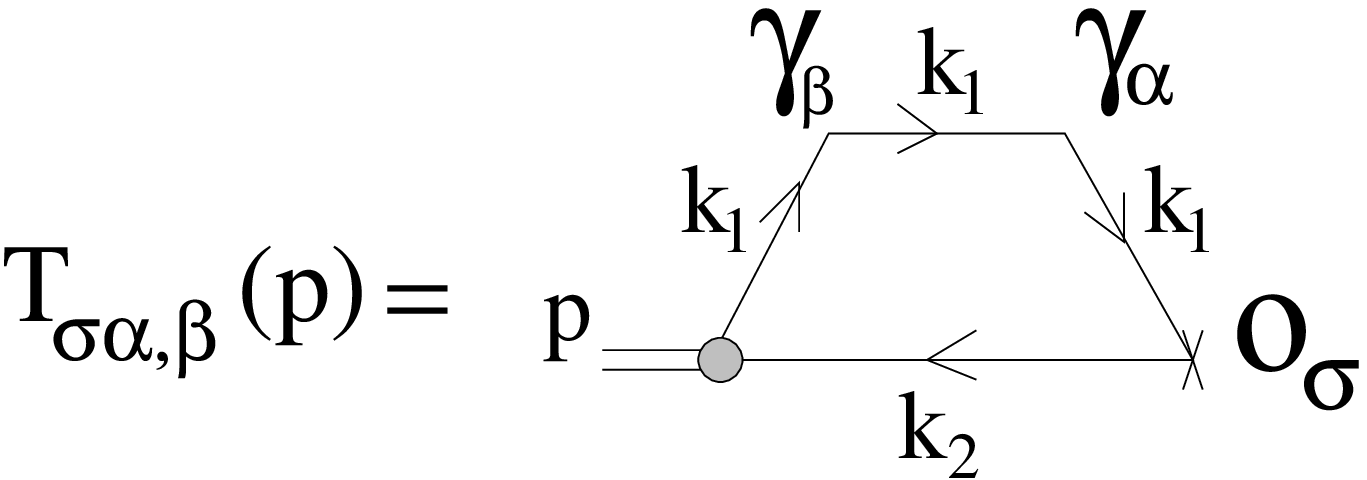,width=7.5cm}}
\end{tabular}
\caption{\label{fig:fig4}
Diagrams for 
$T_{\sigma\alpha}(p,q)$ (a) and 
$T_{\sigma\alpha,\beta}(p)=\frac{\partial }{\partial q^\beta} T_{\sigma\alpha}(p,q)|_{q=0}$
(b). }
\end{center}
\end{figure}

We start with the amplitude $T^{(l)ab}_{\sigma\alpha}(p,q)$. It is diaginal in color
indices, so we write 
$T^{(l)ab}_{\sigma\alpha}(p,q)=\frac{\delta^{ab}}{\sqrt{N_c}}T^{(l)}_{\sigma\alpha}(p,q)$. 
The triangle diagram for $T^{(l)}_{\sigma\alpha}(p,q)$ is shown in Fig. 
\ref{fig:fig4}a. Throughout this section we shall omit the superscript $l$. 
The quark structure of the 
$B$-meson is described by the vertex $i\bar q_1\gamma_5 q_2 G(s)/\sqrt{N_c}$, 
where $s$ is the invariant mass of the $q_1\bar q_2$ pair \cite{m}. One should set 
$m_1=m_b$, $m_2=m_d$ for the calculation of $f^{(b)}$, and $m_1=m_d$, $m_2=m_b$ for $f^{(d)}$. 

We need to calculate $T_{\sigma\alpha,\beta}(p)=
\frac{\partial }{\partial q^\beta} T_{\sigma\alpha}(p,q)|_{q=0}$ (\ref{relation}).  
For the $T_{\sigma\alpha}(p,q)$ shown in Fig \ref{fig:fig4}a, its derivative 
$T_{\sigma\alpha,\beta}(p)$ is given by the diagram in Fig \ref{fig:fig4}b. 
The trace corresponding to the Feynman diagram of Fig \ref{fig:fig4}b reads:   
\begin{eqnarray}
S_{\sigma\alpha;\beta}=
{\rm Sp}\left[ i\gamma_5 (m_2-\hat k_2)O_\sigma (m_1+\hat k_1)
\gamma_\alpha(m_1+\hat k_1)\gamma_\beta(m_1+\hat k_1)\right], 
\end{eqnarray}
where $O_\sigma=\gamma_\sigma(1-\gamma_5)$. Since this expression is further multiplied by 
$G_{\alpha\beta}$, only its antisymmetric part  in $\alpha$ and $\beta$ 
is necessary. The latter reads 
\begin{eqnarray}
S^A_{\sigma\alpha;\beta}&=&(m_1^2-k_1^2)
{\rm Sp} \left(i\gamma_5 (m_2-\hat k_2)\gamma_\sigma(1-\gamma_5) 
\gamma_\beta(m_1-\hat k_1) \right)\nonumber\\
&=&(m_1^2-k_1^2)
\left[4\epsilon_{\beta\alpha\sigma\nu}(m_2 k_1^\nu+m_1 k_2^\nu)
+4im_1 k_2^\nu
(g_{\alpha\sigma}g_{\beta\nu}-g_{\beta\sigma}g_{\alpha\nu})\right].  
\end{eqnarray}
The corresponding expression for the Feynman amplitude takes the form 
\begin{eqnarray}
\label{ta}
T^A_{\sigma\alpha;\beta}(p)&=&\frac{1}{(2\pi)^4i}\int dk_1 dk_2 \delta (p-k_1-k_2)
\frac{-1}{(m_1^2-k_1^2)^2(m_2^2-k_2^2)}
\nonumber\\&&\times
\left[4\epsilon_{\beta\alpha\sigma\nu}(m_2 k_1^\nu+m_1 k_2^\nu)
+4im_1 k_2^\nu(g_{\alpha\sigma}g_{\beta\nu}-g_{\beta\sigma}g_{\alpha\nu})\right]. 
\end{eqnarray}
We define $f_v$ and $f_s$ by the relations 
\begin{eqnarray}
\label{fv}
\frac{1}{(2\pi)^4i}\int dk_1 dk_2\delta (p-k_1-k_2)
\frac{k_1^\nu}{(m_1^2-k_1^2)^2(m_2^2-k_2^2)}&=&p^\nu f_v(p^2),\nonumber\\
\label{fs} 
\frac{1}{(2\pi)^4i}\int dk_1 dk_2\delta (p-k_1-k_2)\frac{1}{(m_1^2-k_1^2)^2(m_2^2-k_2^2)}&=& f_s(p^2). 
\end{eqnarray}
After the $k$-integration in Eq. (\ref{ta}) one recovers the structure of the amplitude from Eq. (\ref{aaa})
\begin{eqnarray}
\label{formula}
T^A_{\sigma\alpha;\beta}(p)=-4p^\nu
\left[f_1 \; \epsilon_{\beta\alpha\sigma\nu} 
+if_2\;(g_{\alpha\sigma}g_{\beta\nu}-g_{\beta\sigma}g_{\alpha\nu})\right],   
\end{eqnarray}
with 
\begin{eqnarray}
f_1&=&(m_2-m_1)f_v+m_1f_s,\nonumber\\
f_2&=& m_1 (f_s-f_v). 
\end{eqnarray}
Let us obtain now spectral representations for the form factors $f^F_v$ and $f^F_s$. 
We first notice that the expressions (\ref{fv}) correspond to the 
triangle Feynman diagram at zero momentum transfer $q=0$. This condition means that both 
$q^2=0$ and $p^2=p'^2$, where $p'=p-q$, so we can write 
$f^F_{i}(p^2)=f^F_{i}(q^2=0,p^2,p'^2=p^2)$, with $i=v,s$.   

It is convenient to start with the case $q^2\ne 0$ and $p^2\ne p'^2$. 
Then the form factors $f_i$ can be written in the form of 
the double spectral representation \cite{m}:  
\begin{eqnarray}
\label{double}
f^F_i(q^2,p^2,p'^2)=\int \frac{ds}{s-p^2}\frac{ds'}{s'-p'^2}\Delta_i (s,s',q^2), 
\end{eqnarray}
where the double spectral density $\Delta_i$ can be calculated for any of the form factors 
from the Feynman integral. 

As the next step, we set $q^2=0$, but still treat $p^2$ and $p'^2$ as independent variables.  
In this case the double spectral representations of the form (\ref{double}) simplify 
to the following single spectral representations 
\begin{eqnarray}
\label{dfs1}
f^F_s(p^2,p'^2)&=&\frac{1}{16\pi^2}
\int ds \frac{1}{s-p^2}\frac{1}{s-p'^2}
\log\left(\frac{s+m_1^2-m_2^2+\lambda^{1/2}(s,m_1^2,m_2^2)}
{s+m_1^2-m_2^2-\lambda^{1/2}(s,m_1^2,m_2^2)}\right),\nonumber\\ 
\label{dfv1}
f^F_v(p^2,p'^2)&=&\frac{1}{16\pi^2}
\int ds \frac{1}{s-p^2}\frac{1}{s-p'^2}
\frac{\lambda^{1/2}(s,m_1^2,m_2^2)}{s}.
\end{eqnarray}
The representations for the form factors in the form (\ref{dfv1}) 
are valid however only in the region of $p^2$ and $p'^2$ (far) below the threshold. 
The reason for that is the following: 
the representations (\ref{double}) and (\ref{dfv1}) are based on the Feynman form of the 
quark propagators. This form is however valid only for highly virtual particles, while in 
the soft region it is strongly distorted by nonperturbative effects. In particular, 
the pole at $k^2=m^2$ in the propagator of a color object, like quark and gluon, is absent. 
Recently, this was confirmed in a lattice study of the gluon propagator \cite{latprop}. 

The modification of the quark propagator in the soft region leads to the 
change of the spectral representation for the form factors in the region 
of $p^2$ and $p'^2$ near and above the $b\bar q$ threshold. 
 
Notice that the quantities $1/(s-p^2)$ and $1/(s'-p'^2)$ in Eq. (\ref{double}) are  
the propagators of the initial and final $b\bar q$ states with virtualities $s$ and $s'$, 
and the squared masses $p^2$ and $p'^2$, respectively. 
The nonperturbative effects which modify the quark and gluon 
propagators in the soft region, modify the propagators of the $b\bar q$ states as well. 

We know the character of these changes in the $p^2$-channel corresponding to the $B$-meson: 
As discussed in \cite{m}, soft interaction of quarks effectively 
replace the factor $1/(s-M_B^2)$ with a regular $B$-meson soft wave function $\phi_B(s)$ in 
the integrand of (\ref{dfv1}).  
The $B$-meson soft wave function is normalized as follows \cite{m}  
\begin{eqnarray}
\label{norm}
\frac{1}{8\pi^2}
\int ds\phi_B^2(s) \frac{\lambda^{1/2}(s,m_b^2,m_q^2)}{2s}(s-(m_b-m_q)^2)=1. 
\end{eqnarray}
The leptonic decay constant is given by the expression 
\begin{eqnarray}
\label{fp}
f_B=
\frac{\sqrt{N_c}(m_b+m_q)}{8\pi^2}
\int ds\phi_B(s) \frac{\lambda^{1/2}(s,m_b^2,m_q^2)}{s}\frac{s-(m_b-m_q)^2}{s}. 
\end{eqnarray}

In the color-octet $p'^2$-channel the Feynman propagator $1/(s-p'^2)$ in Eq (\ref{dfv1}) 
is replaced by the propagator $D(s,p'^2)$ which involves proper modifications in the 
soft region. We therefore find the following representation 
for the form factors $f_i(p'^2)\equiv f_i(q^2=0, p^2=M_B^2, p'^2)$  
\begin{eqnarray}
\label{fvd}
f_v(p'^2)&=&\frac{1}{16\pi^2}
\int ds \phi_B(s)D(s,p'^2)
\frac{\lambda^{1/2}(s,m_1^2,m_2^2)}{s}, 
\nonumber
\\
\label{fsd}
f_s(p'^2)&=&\frac{1}{16\pi^2}
\int ds \phi_B(s)D(s,p'^2)
\log\left(\frac{s+m_1^2-m_2^2+\lambda^{1/2}(s,m_1^2,m_2^2)}
{s+m_1^2-m_2^2-\lambda^{1/2}(s,m_1^2,m_2^2)}\right). 
\end{eqnarray}
We know that $D(s,p'^2)\sim 1/(s-p'^2)$ at large values of $s-p'^2$, and also that 
$D(s,p'^2)$ 
is finite at $s=p'^2$. However, we do not know details of $D(s,p'^2)$ in the 
soft region.   
Motivated by the discussion in \cite{zakharov}, we assume here that the nonperturbative 
effects 
can be described by the following modification of the propagator 
\begin{eqnarray}
\label{prop} 
D(s,p'^2)=\frac{1}{s-p'^2+M_0^2}
\end{eqnarray}
where $M_0$ is a mass parameter. In order to guarantee the absence of the pole 
in $D(s,p'^2)$ in the heavy quark limit, $M_0$ should scale with the heavy quark mass as 
follows: $M_0^2=O(\Lambda_{QCD}\,m_Q)$. So it is convenient to write $M_0$ in the form 
\begin{eqnarray}
\label{m0} 
M_0^2=w\, m_d\,m_b, 
\end{eqnarray}
where 
$m_d\simeq \Lambda_{QCD}$ is the constituent mass of the light quark, and 
$w$ is a parameter of order unity.\footnote{
Assuming that for the color-octet light $q\bar q$ system Eq (\ref{prop}) remains valid with 
$M_0^2=w m_d^2$, we find $D(k^2=0)\simeq 15\; {\rm GeV}^{-2}$ for $w=1$. 
This is close to the value of the gluon propagator $D(k^2=0)\simeq 18\;{\rm  GeV}^{-2}$ 
as found in \cite{latprop}. This agreement seems to be reasonable since one can expect 
the propagator of the light $q\bar q$ color-octet system to have a structure in the 
nonperturbative region similar to the structure of the gluon propagator.} 

In the next section we make use of Eqs. (\ref{fvd}) and (\ref{prop}) to analyse $f_{v,s}(p'^2)$ 
for $B_d$ and $B_s$ mesons. 
\section{Numerical results.}
The parametrization for the $B$-meson wave function 
and values of the quark masses have been determined 
from the analysis of the exclusive processes in \cite{mb,ms}. 
The $B$-meson soft wave function can be written in the form 
\begin{equation}
\label{vertex}
\varphi(s) = \frac{\pi}{\sqrt{2}} \frac{\sqrt{s^2 - (m_q^2 -
m_{\bar{q}}^2)^2}} {\sqrt{s - (m_q - m_{\bar{q}})^2}} \frac{w(k^2)}{s^{3/4}}
\end{equation}
with $k^2=\lambda(s, m_q^2, m_{\bar{q}}^2)/4s$.
As found in \cite{ms} a simple Gaussian parametrization 
$w(k^2)\sim \exp(-k^2/2\beta^2)$ 
gives a good description of the $B$-meson properties. The numerical parameters of the
model from \cite{ms} are listed in Table \ref{table:parameters}. 
\begin{table}[hbt]
\caption{\label{table:parameters}
Constituent quark masses,  slope parameters of the Gaussian wave function, and the 
corresponding calculated leptonic decay constants in GeV from \protect\cite{ms}.}
\centering
\begin{tabular}{|cc|c|c|c|c|c|c|}
& $m_u$ & $m_s$ & $m_b$  & $\beta_B$ &  $\beta_{B_s}$ &  $f_B$  & $f_{B_s}$ \\
& 0.23 & 0.35  &  4.85   & 0.54      &     0.56       &  0.18   &  0.20 
\end{tabular}
\end{table}
The procedure of Ref. \cite{mb,ms} determines the wave function by fitting the lattice 
data on the weak transition form factors for large momentum transfers at the normalization point 
$\mu\simeq 5\;$ GeV. Therefore, the $B$-meson soft wave function and the 
form factors $f_{s,v}(p'^2)$ at this scale are determined. 

\begin{figure}[htb]
\begin{center}
\mbox{\epsfig{file=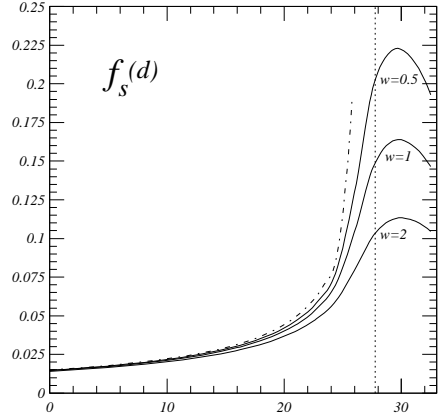,height=6.cm}}
\caption{\label{fig:oscbd}
The form factor $f_s{(p'^2)}$ for the $B_d$ meson vs $p'^2$.  
Solid curves are results of the calculation via the Eqs. \protect(\ref{fvd}) and 
\protect(\ref{prop}) for various values of the parameter $w$: upper curve - $w=0.5$, 
middle curve - $w=1.0$, lower curve - $w=2$. The dash-dotted curve 
in the region $p'^2<(m_b+m_d)^2$ corresponds to $w=0$.  
The vertical dotted line corresponds to $p'^2=M_B^2$. }
\end{center}
\end{figure}

Whereas the $B$-meson wave function is known quite well, good information about the details 
of $D(s,p'^2)$ is lacking. We therefore use the simple Ansatz (\ref{prop}) for $D(s,p'^2)$ 
in the full range of $s$ and $p'^2$ and treat $w$ as a free parameter of order unity.  
We expect that the variation of $w$ in the interval $w=0.5\div 2$ provides reasonable 
error estimates for the form factors and $\Delta B$. 
Notice that the value of $\Delta B$ corresponding to 
$w=1$ agrees favourably with the recent lattice estimates, see Table \ref{table:results}.  
Fig \ref{fig:oscbd} shows the form factor $f_s^{(d)}$ which gives the main 
contribution to $\Delta B_B$. 

The calculated form factors are shown in Table \ref{table:ourresults}. 
The relative magnitudes of these form factors can be easily understood taking into 
account their scaling propertiers in the heavy quark limit: 
$f_s^{(d)}\sim m_b^{-1/2}$, $f_s^{(b)}\sim m_b^{-3/2}$, $f_v^{(b)}/f_s^{(b)}\sim 1$. 

The value of $\Delta B$ is obtained from Eq. (\ref{deltab}) taking 
$\left <\frac{\alpha_s}{\pi}GG\right> = 0.012\; {\rm GeV}^4$ from \cite{svz}.   
Using the Wilson coefficent at the scale $\mu=m_b$ we obtain the  
renorm-invariant factors $\hat B_{B_d}$ and $\hat B_{B_s}$ (for definition see \cite{lat2}) 
listed in Table \ref{table:results}.

\begin{table}[htb]
\centering
\caption{\label{table:ourresults}
Form factors and $\Delta B_{B_q}$ with $q=d$ for $B_d$, and  $q=s$ for $B_s$. The factors 
$f_{1,2}(p'^2)$ are calculated at $p'^2=M^2_{B_d}$ for $B_d$ and at $p'^2=M^2_{B_s}$ for $B_s$.
The errors correspond to the interval $w$=0.5$\div$2.}    
\begin{tabular}{|l|r|r|}
                      &                      $B_d$  &          $B_s$  \\
\hline
$f_v^{(b)}=f_v^{(q)}$ &  $9\pm 3\times10^{-3}$      & $8\pm 3\times10^{-3}$  \\
$f_s^{(b)}$           &  $1.1\pm 0.2 \times10^{-2}$ & $9\pm 3 \times10^{-3}$  \\
$f_s^{(d)}$           &  $1.4\pm 0.5\times10^{-1}$  & $1.0\pm 0.4\times10^{-1}$\\
\hline
$f_1^{(b)}$           &  $7\pm1.5\times10^{-3}$     & $8\pm3\times10^{-3}$  \\
$f_2^{(b)}$           &  $5.5\pm 1.5\times10^{-3}$  & $5.5\pm 2.0\times10^{-3}$ \\
$f_1^{(d)}$           &  $7.5\pm 2.5\times10^{-2}$  & $7.5\pm 2.5\times10^{-2}$ \\
$f_2^{(d)}$           &  $3.5\pm 1.0\times10^{-2}$  & $3.0\pm 1.5\times10^{-2}$ \\
\hline
$\Delta B_{B_q}$      &  $ -0.06\pm0.035$           &   $-0.05\pm0.03$        
\end{tabular}
\end{table}

Our results are in good agreement with the lattice results, 
except for a slightly different estimate of the SU(3) violating effects in the 
$B_d-\bar B_d$ and $B_s-\bar B_s$ cases. This nice agreement gives support to our 
Ansatz for the description of the nonperturbative effects in the propagator of the 
color-octet $q\bar q$ system. 

We would like to notice that taking the Ansatz 
(\ref{prop}) and (\ref{m0}) for the propagator of the $s\bar q$ system 
and setting $M_0^2=w\, m_s\,m_{\bar q}$, with $w\simeq 1$ allows a 
parameter-free estimate of the $K^0-\bar K^0$ mixing. 
In this case the result is much more stable with respect to the particular value of $w$ 
and therefore to the details of the propagator in the nonperturbative region.  
We obtained this way $\Delta B_K(1\; {\rm GeV})=-0.21 \pm 0.04$ \cite{mn}. 
\section{Conclusion.}
We considered corrections to factorization in the $B\bar B$ mixing amplitude due to 
soft-gluon exchanges, assuming that the main effect of such exchanges 
can be described in terms of the local gluon condensate. 

\vspace{.1cm}

\noindent 1. It was demonstrated that within this
approximation correction of the order $\alpha_s\langle GG\rangle$ to the factorization 
is {\it negative}. It can be expressed through the specific $B$-meson transition 
form factors at zero momentum transfer. 

\vspace{.1cm}

\noindent 2. A relativistic dispersion approach based on the constituent quark picture 
has been used to calculate these form factors. Spectral representations for the form factors 
in terms of the $B$-meson soft wave function and the propagator of the color-octet 
$q\bar b$ system were obtained. The behaviour of this propagator in the nonperturbative 
region was discussed. The obtained numerical estimates for $\Delta B_B$ 
favourably compare with the recent lattice calculations.

\vspace{.1cm}

Let us point out that the proposed approach can be extended to the description of more 
complicated problems. In particular, it can be applied to the analysis of the non-factorizable 
effects in non-leptonic $B$ decays. 

\noindent
\acknowledgments
The authors are grateful to H. G. Dosch, O. Nachtmann, and B. Stech for discussions 
and interest in this work. D.M. was supported by the BMBF under project 05 HT 9 HVA3.

\end{document}